# A simple approach of broadband mid-infrared pulse generation with a mode-locked Yb-doped fiber laser

Takuma Nakamura[1], Venkata Ramaiah Badarla[1], Kazuki Hashimoto[1], Peter G. Schunemann[2], Takuro Ideguchi[1]*

1 Institute for Photon Science and Technology, The University of Tokyo, Tokyo 113-0033, Japan
2 BAE Systems, MER15-1813, P.O. Box 868, Nashua, New Hampshire 03061-0868, USA
*Corresponding author: ideguchi@ipst.s.u-tokyo.ac.jp

**Abstract**
Broadband mid-infrared (MIR) molecular spectroscopy demands a bright and broadband light source in the molecular fingerprint region. To this end, intra-pulse difference frequency generation (IDFG) has shown excellent properties among various techniques. However, previous IDFG systems have mainly used unconventional long-wavelength 2-μm ultrashort pulsed lasers. A few systems have been demonstrated with 1-μm lasers, but they use bulky 100-W-class high-power Yb thin-disk lasers. In this work, we demonstrate a simple and robust approach of 1-μm-pumped broadband IDFG with a conventional mode-locked Yb-doped fiber laser. We first generate 3.3-W, 12.1-fs ultrashort pulses at 50 MHz by a simple combination of spectral broadening with a short single-mode fiber and pulse compression with chirped mirrors. Then, we use them for pumping a thin orientation-patterned gallium phosphide (OP-GaP) crystal, generating 1.2-mW broadband MIR pulses with the -20-dB bandwidth of 480 $cm^{-1}$ in the fingerprint region (760-1240 $cm^{-1}$, 8.1-13.1 μm). The 1-μm-based IDFG system allows for simultaneous generation of ultrashort pulses in the ultraviolet and visible regions, enabling, for example, 100-MHz-level high-repetition-rate vibrational sum-frequency generation spectroscopy or pump-probe spectroscopy.

Spectroscopy in the molecular fingerprint region (500-1800 cm$^{-1}$) has a vital role in identifying the composition and structure of molecules in a variety of fields. The non-invasive and label-free optical measurement is essential for the physical, chemical, and biological analysis of molecular substances in various phases. Analyzing multiple molecular species demands a bright and broadband light source in the mid-infrared (MIR) region, where various light sources have been demonstrated [1], including black body radiation, quantum cascade lasers [2], supercontinuum generation [3], optical parametric amplifiers [4], optical parametric oscillators [5,6], and difference frequency generation (DFG) [7]. However, each source has technical challenges in brightness, spectral bandwidth, spectrum flatness, robustness, repetition rate, etc. Among them, intra-pulse DFG (IDFG) with a 100-MHz-level high-repetition-rate laser has shown attractive capabilities for broadband spectroscopy with a flat spectrum covering the fingerprint region and high average power of 100 mW [8]. Since it is based on a mode-locked laser, one can use it as a light source for state-of-the-art spectroscopy techniques such as dual-comb spectroscopy [9], field-resolved spectroscopy [10], and time-stretch spectroscopy [11]. The robustness is another notable aspect of the IDFG source, coming from the single-pass configuration. In addition, DFG from a single laser source inherently has an advantage in producing an offset-free frequency comb [12]. For generating broadband MIR pulses via IDFG, it is beneficial to use a long-wavelength pump source. Therefore, previously demonstrated systems are mainly based on pump lasers at a wavelength above 2 µm [13–17], which are still uncommon devices to date.

Although hindered behind the 2-µm-pumped systems, 1-µm-pumped IDFG-MIR sources are attractive for various reasons. First, the 1-µm-pumped system allows for simultaneous harmonic generations of ultrashort pulses in the ultraviolet (UV) and visible (VIS) regions. The fundamental near-infrared (NIR) 1-µm pulses can also be used simultaneously. The synchronized UV, VIS, NIR, and MIR pulses enable, for example, vibrational sum-frequency generation (VSFG) spectroscopy for surface molecular science [18] and pump-probe spectroscopy for studying ultrafast dynamics of, e.g., photo-reactive proteins such as Rhodopsins [19]. Historically, these experiments require using optical parametric amplifiers based on a Ti:Sapphire regenerative amplifier system running at a low repetition rate of 1 kHz. The 1-µm-pumped IDFG-MIR sources could replace the bulky and inefficient systems. Second, there is a great potential to make the system compact and robust because the technology of 1-µm lasers is well established. In particular, Yb-doped fiber laser is recognized as a gold standard technology in the field of high-power applications such as laser processing [20], and also as a stable frequency comb source for precision measurements [21]. Despite these advantages, there have been a few demonstrations of 1-µm-pumped IDFG-MIR sources except for systems with bulky 100-W-class thin-disk lasers [8,22].

There are three difficulties for making broadband 1-μm-pumped IDFG systems. First, there are limited choices of nonlinear crystals, which must be broadly transparent covering 1 μm and MIR regions. In addition, since IDFG requires 10-fs-level pulses, the two-photon absorption could be a severe problem, making the damage threshold much lower than the case with the 2-μm pump. As a result, only a few crystals, including LiGaS2 (LGS) and orientation-patterned gallium phosphide (OP-GaP), are candidates. Second, a strict phase-matching condition makes the IDFG spectrum narrower because nonlinear crystals have steeper slopes of the refractive index in a shorter wavelength. For example, in our simple calculation of the phase-matching condition in an OP-GaP crystal, more than 5-fold narrower MIR bandwidth is expected than in the case of the 2-μm pump. Third, a considerable group velocity mismatching reduces an interaction length in the crystal—for example, ref.[15] suggests that a 15-fs pulse has an interaction length of 1.5 mm for a 2.5-μm pump in a gallium selenide (GaSe) crystal, while 20 μm for a 1-μm pump. For mitigating these downsides, our strategy is to use a thin OP-GaP crystal and pump it with 10-fs-level ultrashort pulses. The OP-GaP is an excellent quasi-phase-matching crystal with a high nonlinear figure of merit ($d^2/n^3$ >150) [23], a high damage threshold (>0.8 J/cm$^2$) even with a 1064-nm pump [24], and a high thermal conductivity (110 W/mK).

In this work, we demonstrate broadband MIR pulse generation via the IDFG process pumped with a compact 1-μm mode-locked Yb-doped fiber laser at a repetition rate of 50 MHz. First, we generate 12-fs NIR pulses by spectral broadening and pulse compression with a short single-mode fiber (SMF) and chirped mirrors only. Then, we use them for pumping a thin OP-GaP crystal for IDFG. This simple approach generates broadband MIR pulses spectrally covering a wide span of 480 cm$^{-1}$ (760-1240 cm$^{-1}$) and 450 cm$^{-1}$ (570-1020 cm$^{-1}$) with an average power of 1.2 and 0.5 mW, respectively with different grating periods of the OP-GaP crystals. This is the first demonstration of an IDFG-MIR source pumped with a compact Yb-doped fiber laser system to the best of our knowledge. The 100-MHz-level high-repetition-rate IDFG-MIR system could drastically shorten the measurement time of the pump-probe spectroscopy with the UV/VIS/NIR and MIR pulses by a factor of 10$^5$ compared to that with conventional 1-kHz laser systems.

Figure 1 shows a schematic diagram of the developed laser. A 50-MHz mode-locked Yb-doped fiber laser (Menlo systems) is amplified with a homemade ytterbium-doped fiber amplifier (YDFA) and compressed with a grating pair, generating 230-fs pulses with an average power of 5 W. All fiber components are polarization maintained. The pulses are spectrally broadened with a 43-mm standard SMF (HI1060) [25,26] and compressed with chirped mirrors, which compensate for the second-order dispersion by -1300 fs$^2$. This simple scheme results in a robust generation of high-average-power (3.3 W) ultrashort pulses with a pulse duration of 12.1 fs. It is noteworthy that there is no need to make demanding third-order dispersion compensation for the compression. We also note that there has been no optical damage or degradation for months, even though the 5-W, 230-fs pulse (~100 nJ) couples into the 6-μm core of the SMF. Figure 2 shows the generated broadened spectrum measured with a homemade Fourier-transform spectrometer. The spectral broadening is mainly due to the self-phase modulation, which is well-predicted by a simple nonlinear Schrödinger equation (NLSE). Figure 3(a) shows an autocorrelation trace of the compressed pulses measured with a homemade auto-correlator, showing the 12.1 fs pulse duration with an assumption of hyperbolic secant squared (sech2) function. Thanks to the well-known optical property of the standard HI1060 fiber [25], one can predict the amount of negative dispersion for compensating the dispersion of the SMF within the error of less than 100 fs$^2$ by the NLSE (Fig. 3(b)), making the system design simple. It is worth mentioning that our approach to generating 10-fs pulses with an SMF and chirped mirrors is much simpler and more robust than the previously demonstrated other schemes with a photonic crystal fiber [8], a Herriott-type multi-pass cell [27], a large mode area fiber [22], a highly nonlinear fiber [28,29], an anti-resonant hollow-core fiber [13], nonlinear $\chi^{(3)}$ interactions in the amplifier's gain element [30], etc.

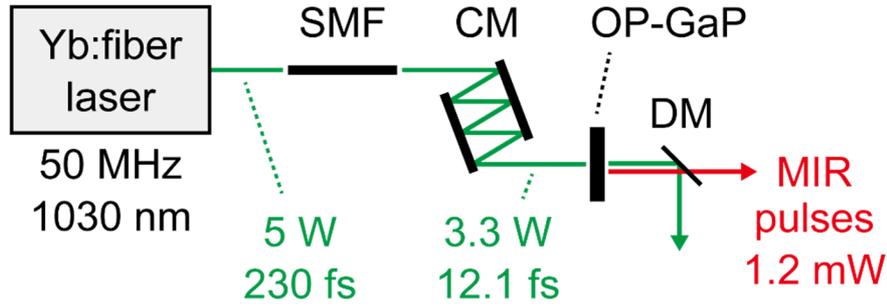

**Fig. 1. Schematic of the 1-μm-pumped IDFG system based on a Yb-doped fiber laser.**
A single-mode fiber broadens the spectrum, and chirped mirrors compress the pulses, generating 3.3-W, 12.1-fs ultrashort pulses. They are focused onto an orientation-patterned gallium phosphide crystal, generating 1.2-mW broadband MIR pulses. SMF, single-mode fiber; CM, chirped mirror; OP-GaP, orientation-patterned gallium phosphide; DM, dichroic mirror; MIR, mid-infrared.

For generating MIR pulses by IDFG, the 3.3-W, 12.1-fs pulses are focused onto a 0.8-mm-long OP-GaP crystal (side 1 and 2 anti-reflection (AR) -coated at 1040-1080 nm, 1150-1350 nm, and 5000-12000 nm) with a grating period of 27 μm. The input pulses are negatively pre-chirped to compensate for a large dispersion of the OP-GaP crystal at 1030 nm (>1300 fs$^2$/mm [31]). We use off-axis parabolic mirrors to focus and collimate the ultrashort pulses. A custom-made ZnSe dichroic mirror (DM) separates the undepleted high-power 1-μm pump pulses after the OP-GaP crystal, which can be used for various applications such as electro-optic sampling [8,32], up-conversion spectroscopy [33], complementary vibrational spectroscopy [34], and UV/VIS harmonic generations. Then, a 4.5-μm Germanium long-pass filter removes weak residual pump pulses leaking the DM. The average power of the extracted MIR pulses is 1.2 mW, which is high

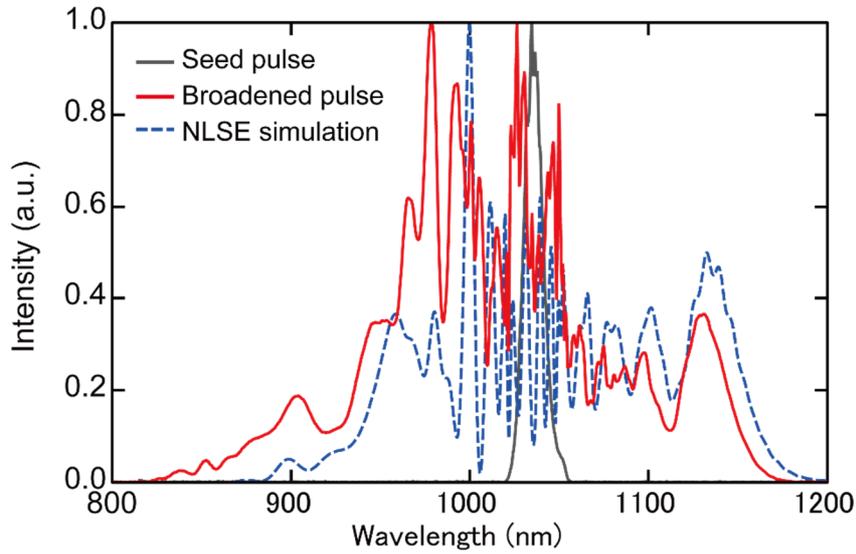

**Fig. 2. A spectrum of the 1-μm pulses broadened with the 43-mm SMF.**

The black line shows a spectrum of the seed pulses before the SMF. The blue dashed line shows a simulated spectrum calculated by NLSE.

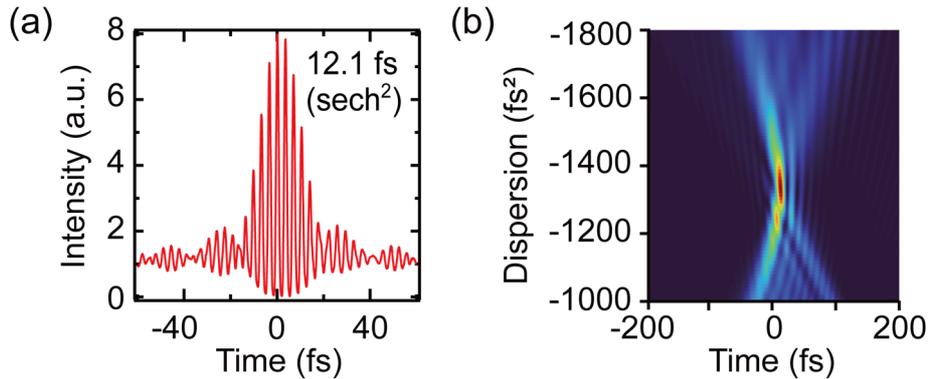

**Fig. 3. Temporal characterization of the compressed 1-μm pulses.**

(a) An autocorrelation trace of the compressed 1-μm ultrashort pulses, showing a pulse duration of 12.1 fs with an assumption of hyperbolic secant squared (Sech$^2$) temporal waveform. (b) A simulated temporal intensity profile of a pulse after the SMF as a function of the additional dispersion. The shortest pulse duration is predicted at -1350 fs$^2$.

enough for most spectroscopy applications because about 90% power loss is still acceptable to use up a full dynamic range of a typical HgCdTe (MCT) detector. The red curves in Fig. 4 (a: linear scale, b: logarithmic scale) show a spectrum of the IDFG-MIR pulses measured with a homemade Fourier transform infrared (FTIR) spectrometer with an MCT detector whose cut-on wavenumber is around 500 cm$^{-1}$ (20 μm). The -20-dB spectral bandwidth is 480 cm$^{-1}$ spanning 760-1240 cm$^{-1}$ (8.1-13.1 μm), broadly covering the fingerprint region. The spectral range can shift

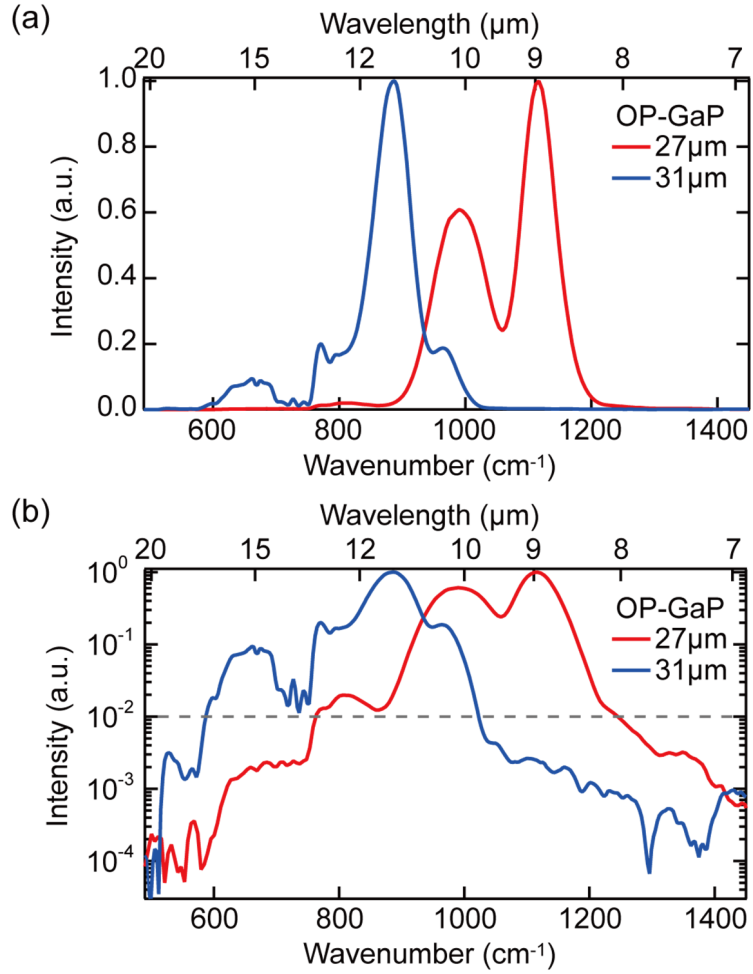

**Fig. 4. Broadband mid-infrared spectra generated via 1-μm pumped intra-pulse difference frequency generation with OP-GaP crystals.**

(a) Spectra at linear scale. (b) Spectra at logarithmic scale. The gray dashed line represents a -20 dB level.

with a different grating period of the OP-GaP crystal. The blue curves in Fig.3 show a spectrum generated with a grating period of 31 μm (same AR coating), which covers 590-1020 cm$^{-1}$ (9.8-17.0 μm) at a -20-dB attenuation level. The large dip around 730 cm$^{-1}$ in the spectrum is attributed to the absorption of the OP-GaP crystal itself, resulting in a slightly lower IDFG power of 0.5 mW.

In this proof-of-concept demonstration, the OP-GaP crystal was not perfectly optimized for our pump source. By fully utilizing the broadband and well-compressed pump pulses spanning 820-1170 nm, it is possible to generate an enlarged spectrum toward a higher wavenumber region with an optimum grating period. For example, IDFG centered at 1660 cm$^{-1}$ (6 μm) is expected with a grating period of 19 μm. A fan-out grating [32] could provide a widely tunable source in the fingerprint region (500-1800 cm$^{-1}$), while a chirped grating could generate an ultra-broadband spectrum in this region. A periodically-polled lithium niobate (PPLN) crystal could provide IDFG at an even higher-wavenumber region around 3000 cm$^{-1}$. In addition, an optimum AR coating on the surface of the OP-GaP crystal is important to mitigate a large Fresnel loss due to a high refractive index (>3). In our experiment, the AR coating of the crystal does not support the region below 1040 nm for the pump pulses, causing the undesired reduction of the spectral bandwidth. For example, in our case, for generating a MIR spectrum around 800 cm$^{-1}$ with the 27-μm-period crystal, the pump pulses must contain a spectral component around 910 nm, which suffers from the large Fresnel loss due to the imperfect AR coating. Therefore, an optimum AR coating could improve the conversion efficiency and expand the spectral bandwidth.

It is helpful to compare our laser to another fiber-laser-based IDFG source, particularly an OP-GaP-based IDFG-MIR laser pumped by an Er-doped fiber laser [28]. The main difference comes from the pump wavelengths centered at 1030 nm and 1550 nm, respectively. A simple calculation tells us the 1-μm-pumped OP-GaP-IDFG has disadvantages in 3-fold narrower spectral bandwidth due to the steeper phase slope and 5-fold shorter interaction length due to the larger group velocity mismatch. On the other hand, our 1-μm-pumped system generates more than 4-fold higher power than that of the 1.5-μm-pumped counterpart due to the inherently larger amplification capability of YDFA than erbium-doped fiber amplifier (EDFA). Another advantage of the 1-μm-pumped system is accessibility to the UV/VIS region, enabling various spectroscopy techniques that use both electronic transitions in the UV/VIS and nuclear vibrations in the MIR.

In conclusion, we demonstrated an IDFG-MIR source with a thin OP-GaP crystal pumped with a compact 1-μm mode-locked Yb-doped fiber laser at a repetition rate of 50 MHz. The developed laser provided mW-level high-power and broadband MIR pulses with the -20-dB bandwidth of 480 cm$^{-1}$ spanning 760-1240 cm$^{-1}$, and 590-1020 cm$^{-1}$ with different OP-GaP crystals' grating periods. The 1-μm-pumped high-repetition-rate high-power IDFG-MIR laser would be useful in applications requiring ultrashort pulses in both the UV/VIS and MIR regions, such as VSFG or pump-probe spectroscopy. The 100-MHz-level high-repetition-rate laser system has the potential to replace the conventional kHz-level low-repetition-rate Ti:Sapphire regenerative amplifier laser systems, enabling various ultrafast measurements in an orders-of-magnitude faster or more sensitive manner.

**Funding.**

Japan Society for the Promotion of Science (20H00125); Precise Measurement Technology Promotion Foundation (PMTP-F); JST PRESTO (JPMJPR17G2); Japan Society for the Promotion of Science (21K20500).

**Acknowledgments.**

The authors thank Satoko Yagi for the help in making the FTIR spectrometer and autocorrelator systems.

**Disclosures.**

The authors declare no competing interests.

**Data availability.**

The data that support the findings of this study are available from the corresponding author upon reasonable request.